\documentclass[a4paper,twocolumn,showpacs,aps,prd,nofootinbib]{revtex4}
\usepackage[latin9]{inputenc}
\usepackage{hyperref}
\hypersetup{
  colorlinks=true,        
  linkcolor=blue,         
  citecolor=cyan,         
}
\usepackage{breakurl}
\usepackage{graphicx}
\usepackage{epsf}
\usepackage{epsfig}
\usepackage{amssymb,amsmath}
\usepackage[usenames]{color}
\usepackage{amssymb}
\usepackage{times}

\newcommand{\cf}{cf.~}
\newcommand{\ie}{i.e.,~}
\newcommand{\eg}{e.g.,~}



\begin{document}


\title{Analytic Bjorken flow in one-dimensional relativistic
  magnetohydrodynamics}

\author{Victor Roy$^{1}$, Shi Pu$^{1}$, Luciano Rezzolla$^{1,2}$, Dirk Rischke$^{1}$}

\affiliation{$^{1}$ Institute for Theoretical Physics, Goethe University, 
Max-von-Laue-Str.\ 1, 60438 Frankfurt am Main, Germany} 
\affiliation{$^{2}$ Frankfurt Institute for Advanced Studies, 
Ruth-Moufang-Str.\ 1, 60438 Frankfurt am Main, Germany}

\begin{abstract}
In the initial stage of relativistic heavy-ion collisions, strong
magnetic fields appear due to the large velocity of the colliding
charges.  The evolution of these fields appears as a novel and intriguing
feature in the fluid-dynamical description of heavy-ion collisions.  In
this work, we study analytically the one-dimensional, longitudinally
boost-invariant motion of an ideal fluid in the presence of a transverse
magnetic field. Interestingly, we find that, in the limit of ideal
magnetohydrodynamics, \ie for infinite conductivity, and irrespective
of the strength of the initial magnetization, the decay of the fluid
energy density $e$ with proper time $\tau$ is the same as for the
time-honored ``Bjorken flow'' without magnetic field. Furthermore, when
the magnetic field is assumed to decay $\sim \tau^{-a}$, where $a$ is an
arbitrary number, two classes of analytic solutions can be found
depending on whether $a$ is larger or smaller than one. In summary, the
analytic solutions presented here highlight that the Bjorken flow is far
more general than formerly thought.  These solutions can serve both to
gain insight on the dynamics of heavy-ion collisions in the presence of
strong magnetic fields and as testbeds for numerical codes.
\end{abstract}


\maketitle

\section{Introduction}

Very intense magnetic fields of the order of $B\sim10^{18} -10^{19}$ G
are produced orthogonal to the direction of motion in a typical
non-central Au-Au collision at top RHIC energy (\ie with a
centre-of-momentum energy per nucleon pair of $\sqrt{s_{\rm NN}} \simeq
200\,{\rm GeV})$. Recent studies show that the strength of the produced
magnetic field grows approximately linearly with the centre-of-momentum
energy of the colliding nucleons \cite{Bzdak:2011yy, Deng:2012pc,
  Kharzeev:2007jp}. It is now an experimentally well-established fact
that in high-energy nucleus-nucleus collisions a very hot and dense phase
of nuclear matter composed of quarks and gluons is formed. This hot and
dense form of nuclear matter is also known as quark-gluon plasma
(QGP). In the presence of a strong magnetic field as created in heavy-ion
collisions, a charge current will be induced in the QGP, leading to what
is known as the ``chiral magnetic effect'' (CME)
\cite{Kharzeev:2007jp}. At the same time, particles with the same charge
but different chirality will also be separated, yielding what is called
the ``chiral separation effect'' (CSE). A density wave induced by these
two effects, called the ``chiral magnetic wave'' \cite{Kharzeev:2010gd},
is suggested to break the degeneracy between the elliptic flows of
positive and negative pions \cite{Burnier:2011bf}. Moreover, it has also
been found that there exists a deep connection between these effects and
the Berry phase in condensed matter
\cite{Stephanov:2012ki,Son:2012zy,Chen:2012ca}. Research on these topics
is developing rapidly and a series of recent reviews and references can
be found in Refs.\ \cite{Bzdak:2012ia, Kharzeev:2013ffa,
  Kharzeev:2015kna}.

The initial electric fields are also found to be quite large in
event-by-event simulations of heavy-ion collisions
\cite{Deng:2012pc}. Such electric fields induce other novel effects such
as the ``chiral electric separation effect'' (CESE) and the ``chiral
electric wave'' (CEW), which represent chiral currents and density waves
induced by the electric fields, respectively
\cite{Huang:2013iia,Pu:2014cwa}. When the electric field is perpendicular
to the magnetic field, just like for the Hall effect, a chiral current is
expected, called the ``chiral Hall separation effect'' (CHSE), which
might cause an asymmetric charge distribution in rapidity
\cite{Pu:2014fva}.

Relativistic hydrodynamics has been proven to be quite successful in
describing the experimentally measured azimuthal distribution of particle
emission in non-central nucleus-nucleus collisions \cite{Romatschke,
  Heinz, Roy:2012jb, Heinz:2011kt, Niemi:2012ry, Schenke:2011bn}. It is
then important to understand the effect of initial large magnetic fields
on the fluid evolution. To this scope one needs a numerical code that
solves the equations of (3+1)--dimensional relativistic
magnetohydrodynamics (MHD).  There is consensus that due to very high
velocities of the charges inside the colliding nuclei (\ie with Lorentz
factors $\gamma \sim 100$ for collisions at $\sqrt{s_{\rm NN}}=200\, {\rm
  GeV}$), the magnetic fields decay very rapidly (\ie decreasing by
$\sim$ 3 orders of magnitude within a timescale $\sim 1\,{\rm fm}$ for
$\sqrt{s_{\rm NN}}=200\, {\rm GeV}$ Au-Au collisions) before the system
reaches local thermal equilibrium and the fluid description is
applicable. However, the presence of a medium with finite electrical
conductivity can substantially delay the decay of the magnetic field
\cite{Gursoy:2014aka,Zakharov:2014dia, Tuchin:2013apa}.

Lattice-QCD simulations and theoretical models show that the QGP
possesses a finite temperature-dependent electrical conductivity
\cite{Gupta:2003zh,Qin:2013aaa}. However, the interaction of the initial
magnetic field with the QGP and its subsequent evolution is still an open
issue and a topic of current research. An estimate of the relative
importance of an external magnetic field on the fluid evolution can be
obtained from the dimensionless quantity $\sigma \equiv {B^{2}}/{e}$,
which represents the ratio of the magnetic-field energy density to the
fluid energy density $e$. Clearly, values of $\sigma \gtrsim 1$ indicate
that one must consider the effect of the magnetic field on the fluid
evolution. For a typical mid-central (\ie with impact parameter $b\sim
10\,{\rm fm}$) Au-Au collision at top RHIC energy ($\sqrt{s_{\rm NN}} =
200\,{\rm GeV}$) the average magnetic field can be as high as $\sim 10\,
m_{\pi}^{2}\sim10^{19}$ G \cite{Bzdak:2011yy,Deng:2012pc}, where $m_\pi$
is the pion mass, which corresponds to an energy density of $\sim 5\,{\rm
  GeV}/{\rm fm}^{3}$. Hydrodynamical model studies show that the initial
energy density for such cases is $\sim 10\,{\rm GeV}/{\rm fm}^{3}$, thus
implying $\sigma \sim 1$ under these conditions.

We note that the estimates made above are based on the assumption that
the magnetic field (evaluated at time $\tau=0\,{\rm fm}$) remains
unchanged until the fluid starts expanding after reaching local thermal
equilibrium at $\tau_{0}\sim 0.5\, {\rm fm}$. We also note that the
estimate of $\sigma$ as given above is based on the event-averaged values
for the initial magnetic field and energy density of the fluid. However,
the situation can be very different. In fact, it is possible that the
initial energy density distribution is very ``lumpy".  Under these
conditions, the produced magnetic fields also show large variations and
can be very large in some places where the corresponding fluid
energy-density is small. In these cases, even for a quickly decaying
initial magnetic field we may locally have $\sigma > 1$ up to the time
when the hydrodynamical expansion starts.

It is not the goal of this work to investigate the temporal evolution of
the magnetic field produced in heavy-ion collisions. Rather, we
concentrate here on the special case of one-dimensional, longitudinally
boost-invariant fluid expansion \`{a} la Bjorken \cite{Bjorken:1982qr}
under the influence of an external magnetic field which is transverse to
the fluid velocity.  In our analysis the evolution of the magnetic field
is either regulated from the flux-freezing condition in ideal MHD or
imposed in terms of a parameterized power law in proper time. The
ultimate goal is that of finding analytic solutions for this flow that
can be used both to gain insight in the dynamics of ultrarelativistic MHD
flows as well as an effective test for more complex and realistic
numerical codes (see also Ref.\ \cite{Lyutikov:2011vc} for a work with
similar intentions).

The paper is organised as follows.  Section \ref{sec:mathsetup}
introduces our mathematical setup, while Sec.\ \ref{sec:energy_evolution}
presents the energy-density evolution when considering two representative
prescriptions for the evolution of the magnetic field. A discussion of
our main results is presented in Sec.\ \ref{sec:results}, while a
conclusive summary is given in Sec.\ \ref{sec:summary}.

Following the predominant convention in relativistic hydrodynamics of
heavy-ion collisions, we use a timelike signature $(+,-,-,-)$ 
and a system of units in which $\hbar=c=k_{_{\rm B}}=1$. Greek indices are
taken to run from 0 to 3, Latin indices from 1 to 3 and we adopt the
standard convention for the summation over repeated indices. Finally, we
indicate three-vectors as bold face letter with an arrow and use bold letters
without an arrow to denote four-vectors and tensors.

\section{Mathematical setup}
\label{sec:mathsetup}

We consider an ideal but magnetised relativistic fluid with an
energy-momentum tensor given by\footnote{Note that expression
    \eqref{eq:EMTensor} for the energy-momentum tensor is different from
    the one usually adopted in general-relativistic MHD (GRMHD)
    formulations and that we briefly review in App. \ref{Appendix2}.}
\cite{Gedalin:1995,Huang:2011dc,Giacomazzo:2005jy,Giacomazzo:2007ti}
\begin{eqnarray}
\label{eq:EMTensor}
{T}^{\mu\nu}=\left(e+p+{B}^{2}\right){u}^{\mu}{u}^{\nu}-
\left(p+\frac{{B}^{2}}{2}\right){g}^{\mu\nu}-{B}^{\mu}{B}^{\nu}\,,
\end{eqnarray}
where $e, p$, and $\boldsymbol{u}$ are the fluid energy density,
pressure, and four-velocity, respectively. Since our considerations are
restricted to special-relativistic flows, the metric tensor is that of
flat spacetime, \ie ${g}^{\mu\nu} = {\eta}^{\mu\nu} =
\rm{diag}(1,-1,-1,-1)$. Here $B^{\mu}=\frac{1}{2}
\epsilon^{\mu\nu\alpha\beta}F_{\nu\alpha}u_{\beta}$ is the magnetic field
in the frame moving with the velocity $u_{\beta}$,
$\epsilon^{\mu\nu\alpha\beta}$ is the completely antisymmetric four
tensor, $\epsilon^{0123}=\sqrt{\det|g|}$. The magnetic field
  four-vector $B^{\mu}$ is a spacelike vector with modulus
  $B^{\mu}B_{\mu}=-B^{2}$, and orthogonal to $u^{\mu}$, \ie
  $B^{\mu}u_{\mu}=0$, where $B=|\vec{\boldsymbol{B}}|$ and
  $\vec{\boldsymbol{B}}$ is the magnetic field three-vector in the frame
  moving with four-velocity $u^{\mu}$.

As mentioned above, we are here interested in obtaining analytic
solutions representing the MHD extension of one-dimensional Bjorken flow
along the $z$--direction with velocity $u^{\mu}
=\gamma\left(1,0,0,v_z\right)$ , where $v_{z} \equiv z/t$
\cite{Bjorken:1982qr}. Hence, hereafter we will assume the special case
of a fluid flow in which the external magnetic field
$\vec{\boldsymbol{B}}$ is directed along the direction transverse to the
fluid velocity $\vec{\boldsymbol{v}}$; as remarked above, this represents
a rather good approximation of what happens in a typical non-central
Au-Au collision at top RHIC energy. This setup is also known as
``transverse MHD'', since the magnetic field is contained in the
transverse $(x,y)$ plane \cite{Romero:2005ct}. In addition, since the
fluid is expected to be ultrarelativistic, the rest-mass contributions to
the equation of state (EOS) can be neglected and the pressure is simply
proportional to the energy density, \ie
\begin{equation}
\label{eq:EOS}
p={c}_{s}^{2}\, e = \frac{1}{3}\,e\,, 
\end{equation}
where $c_s$ is the local sound speed which is assumed to be constant. The
second equality in Eq.\ \eqref{eq:EOS} refers to the case of an
ultrarelativistic gas, or isotropic ``radiation fluid''
\cite{Rezzolla:2013}, where $c_s=1/\sqrt{3}$ and which we will often
consider in the remainder of this work.

Rather than using a standard Cartesian coordinate system $(t,x,y,z)$, for
longitudinally boost-invariant flow it is more convenient to adopt Milne
coordinates,
\begin{equation}
(\tau,x,y,\eta) \equiv
\left(\sqrt{t^{2}-z^{2}},x,y,
\frac{1}{2}\ln\left(\frac{t+z}{t-z}\right)\right)\,. 
\end{equation}
In these coordinates, the convective derivative is defined as
$u^{\mu}\partial_{\mu} = \partial_{\tau}$, while the expansion scalar
takes the simple form $\Theta \equiv \partial_{\mu}u^{\mu} = \tau^{-1}$.

As customary, the projection of the energy-momentum conservation equation
${\partial}_{\nu}{T}^{\mu\nu}=0$ along the fluid four-velocity,
\begin{eqnarray}
{u}_{\mu}{\partial}_{\nu}{T}^{\mu\nu} & = & 0\,,
\end{eqnarray}
will express the conservation of energy. After some steps that can be
found in App.\ \ref{Appendix2}, we obtain the following
energy-conservation equation
\begin{eqnarray}
\label{eq:1dMHD}
\partial_{\tau} \left(e+\frac{{B}^{2}}{2}\right)
+\frac{e+p+{B}^{2}}{\tau}=0 \,.
\end{eqnarray}
Similarly, the projection of the energy-momentum conservation
equation onto the direction orthogonal to $\boldsymbol{u}$,
\begin{equation}
(\eta_{\mu\nu}-u_{\mu}u_{\nu})\partial_{\alpha}T^{\alpha\nu}=0\,.
\end{equation}
leads to the momentum-conservation, or Euler, equation (see
App.\ \ref{Appendix2}),
\begin{equation}
\label{eq:Euler}
\left(e+p+{B}^{2}\right) \partial_{\tau}\,{{u}_{\mu}}-
(\eta_{\mu\nu} - u_{\mu} u_{\nu})\partial^{\nu}
\left(p+\frac{{B}^{2}}{2}\right)=0\,.
\end{equation}
Note that for $\mu=\eta$, it reads
\begin{equation}
\frac{\partial}{\partial\eta}\left(p+\frac{1}{2}B^{2}\right)=0\,,
\end{equation}
thus showing that all thermodynamical variables depend only on $\tau$ and are
otherwise uniform in space. Considering instead $\mu=x,y$,
Eq.\ \eqref{eq:Euler} gives the MHD equivalent of the Euler equation
\begin{equation}
\label{eq:Euler_1}
\partial_{\tau}u_{i}-\frac{1}{e+p+B^{2}}\partial_{i}
\left(p+\frac{1}{2}B^{2}\right)=0\,.
\end{equation}
With a uniform pressure and a magnetic field that depends only on $\tau$,
the second term in Eq.\ \eqref{eq:Euler_1} will vanish, thus implying that
if the velocities in the $x$-- and $y$--directions are initially zero, they
will remain so also at later times (\ie $\partial_{\tau} u_i = 0$).

\section{Energy-density evolution}
\label{sec:energy_evolution}

This section is dedicated to the discussion of two different cases for
the evolution of the energy density depending on whether the magnetic
field evolves according to the ideal-MHD limit (Sec.\ \ref{sec:IMHD})
or whether it follows an arbitrary power-law decay in proper time (Sec.\
\ref{sec:powerlaw}).

\subsection{Ideal-MHD limit}
\label{sec:IMHD}

The solution of Eq.\ \eqref{eq:1dMHD} requires the knowledge of the
evolution of the magnetic field and hence of the induction equation. In
the limit of infinite electrical conductivity, \ie in the ideal-MHD
limit, the magnetic field obeys the frozen-flux (or Alfv\'en) theorem
and is thus simply advected with the fluid. In this case,
setting $B\equiv \sqrt{B^i B_i}$, the induction equation takes the simple
form
\begin{equation}
\label{eq:fluxfreeze}
B(\tau)=B_0\frac{\rho}{\rho_{0}}\,,
\end{equation}
where $\tau_{0}$ is taken to mark the beginning of the fluid expansion
and $\rho_{0} \equiv \rho(\tau_0)$, $B_{0} \equiv B(\tau_0)$,
are the initial fluid rest-mass density and magnetic field, respectively.

As written, Eq.\ \eqref{eq:fluxfreeze} is of little use. In relativistic
heavy-ion collisions, in fact, the net-baryon number is vanishingly small
at mid-rapidity and the flux-freezing condition expressed by
Eq.\ \eqref{eq:fluxfreeze} needs to be modified to account for this. As
we show in App.\ \ref{appen:frozenFlux}, this is rather easy to do and
yields, for an ultrarelativistic fluid with EOS \eqref{eq:EOS}, an
evolution equation for the magnetic field,
\begin{equation}
\label{eq:BasEntro}
\vec{\boldsymbol{B}}(\tau)=\vec{\boldsymbol{B}}_0\frac{s}{s_{0}}=
\vec{\boldsymbol{B}}_0{\left(\frac{e}{e_{0}}\right)}^{1/(1+c^2_s)}=
\vec{\boldsymbol{B}}_0{\left(\frac{e}{e_{0}}\right)}^{3/4}\,,
\end{equation}
where $s$ is the entropy density, $s_{0} \equiv s(\tau_0)$, and the
third equality in Eq.\ \eqref{eq:BasEntro} refers to the case
$c_s=1/\sqrt{3}$. Note that the second equality in
Eq.\ \eqref{eq:BasEntro} is the result of the first law of thermodynamics
and reflects the relation between entropy and energy densities in an
ultrarelativistic fluid \cite{Rezzolla:2013}.

Using Eq.\ (\ref{eq:BasEntro}) in Eq.\ \eqref{eq:1dMHD}, we obtain
\begin{equation}
\label{eq:1dMHD_2}
\partial_{\tau}\!\left[e\!+\!
\frac{{B}_{0}^{2}}{2} \!\left(\frac{e}{e_{0}}\right)^{2/(1+c^2_s)} \right]\!
      + \frac{e + p + B_{0}^{2}\left(e/e_{0}\right)^{2/(1+c^2_s)}}{\tau} =0\,,
\end{equation}
which can also be written in dimensionless form as
\begin{equation}
\partial_{\tau} \left[ \tilde{e} +\frac{\sigma_{0}}{2} {\tilde{e}}^{2/(1+c^2_s)}
\right] +\frac{\tilde{e} +\tilde{p} +
\sigma_{0}{\tilde{e}}^{2/(1+c^2_s)}}{\tau} =0\,,
\label{eq:1dMHD_3}
\end{equation}
where ${\tilde{e}} \equiv {e}/{{e}_{0}}$, ${\tilde{p}} \equiv
{p}/{{e}_{0}}$, and $\sigma_0 \equiv {{B}_{0}^{2}}/{{e}_{0}}$.

Not surprisingly, in the limit of vanishing magnetic field, \ie for
$\sigma_0 \rightarrow 0$, Eq.~(\ref{eq:1dMHD_3}) takes the form of
the Bjorken expansion, for which the energy density evolves according to
\begin{equation}
\partial_{\tau} \tilde{e} = -
\frac{ \tilde{e} + \tilde{p} }{\tau} =-
(1+c^2_s)\frac{\tilde{e}}{\tau}
\,,
\label{eq:1dBjorken}
\end{equation}
where the second equality is written using the EOS \eqref{eq:EOS}.




Using again Eq.\ \eqref{eq:EOS}, we find that both
terms on the left-hand side of Eq.\ (\ref{eq:1dMHD_3}) contain a common
factor that can removed: $1 + \sigma_0 \tilde{e}^{(1 - c_s^2)/(1 +
  c_s^2)}/(1 + c_s^2)$. As a result, Eq.\ (\ref{eq:1dMHD_3}) can be
written as
\begin{equation}
\label{eq:1dMHD_4}
\partial_{\tau}\, \tilde{e} =-
(1+c^2_s)\frac{\tilde{e}}{\tau} = - \frac{4}{3}
\frac{{\tilde{e}}}{\tau} \,,
\end{equation}
where the second equality refers to the more specific case of an EOS with
$c_s=1/\sqrt{3}$. 

Thus, Eq.\ (\ref{eq:1dMHD_4}), which was derived within ideal MHD,
coincides with Eq.\ (\ref{eq:1dBjorken}), which instead refers to Bjorken
flow in the absence of an external magnetic field.
After a more careful look, this result is not so surprising. In the
ideal-MHD limit, in fact, the ratios $\vec{\boldsymbol{B}}/s$ and thus
$\vec{\boldsymbol{B}}/e^{1/(1+c^2_s)}$ are conserved [\cf
  Eq.\ (\ref{eq:BasEntro})] and although the total energy density will be
larger in the presence of a magnetic field, the evolution of the fluid
energy density will not be affected by the magnetic field which will be
equally diluted as the fluid expansion takes place. This is essentially
because the magnetic field has no active role in ideal MHD, but is simply
passively advected in the expansion. Stated differently, a Bjorken flow
is more general than formerly thought, as it applies not only to purely
hydrodynamical flows, but also to transverse MHD flows. To the best of
our knowledge, this result, albeit natural, was not remarked before in
the literature.

Equation (\ref{eq:1dMHD_4}) has an analytic solution of the form
\begin{equation}
\label{eq:analytic_withB_s}
\tilde{e}(\tau) = c{\tau}^{-1/(1+c^2_s)} = 
\left(\frac{\tau_0}{\tau}\right)^{1+c^2_s}=
\left(\frac{\tau_0}{\tau}\right)^{4/3}\,,
\end{equation}
where $c$ is a constant that could be chosen, for instance, from the
initial value of the energy density: $\tilde{e}_0 \equiv
\tilde{e}(\tau_0) = 1$, and where the last equality again refers to
$c_s=1/\sqrt{3}$. In the light of the remarks made above, it follows that
Eq.\ \eqref{eq:analytic_withB_s} is also the solution for the energy
density for Bjorken flow without magnetic field, \cf
Eq.\ (\ref{eq:1dBjorken}).  

Two final remarks: first, we note that
combining Eqs.\ \eqref{eq:BasEntro} and \eqref{eq:analytic_withB_s}, it
is easy to see that the evolution of the magnetic field in this case
will be
\begin{equation}
\label{eq:Bs_evol}
\tilde{B}(\tau) \equiv \frac{B(\tau)}{B_0} = \tilde{e}^{1/(1+c^2_s)} =
\tilde{e}^{3/4} = \frac{\tau_0}{\tau}\,.
\end{equation}
Second, our conclusion that the Bjorken flow is recovered in transverse
MHD could have been reached also using entropy conservation and the
Maxwell equations as long as 
the thermodynamical relations are not affected by the presence of a
magnetic field (zero magnetization vector), \ie as long as $de = T ds$,
where $T$ is the temperature \cite{Huang:2010sa}. In this case the
derivation does not even require the specification of an EOS.

\subsection{Power-law decay}
\label{sec:powerlaw}

Next, we explore cases where the external magnetic field does not vary
according to the ideal-MHD flux-freezing condition (\ref{eq:BasEntro}),
but has a different temporal evolution. Because we are in search of
analytic solutions, we consider here a rather simple prescription and, in
particular, one in which the magnetic field follows a power-law decay in
proper time, \ie
\begin{equation}
\label{eq:BasTau}
\vec{\boldsymbol{B}}(\tau)=\vec{\boldsymbol{B}}_0
{\left(\frac{\tau_{0}}{\tau}\right)}^{a}\,,
\end{equation}
where $a$ is a constant. Clearly, expression \eqref{eq:BasTau} is a
simple ansatz but, as remarked in Eq.\  \eqref{eq:Bs_evol}, it is
sufficiently realistic to include the ideal-MHD case when $a=1$. In
addition, the range $a > 1$, \ie of magnetic-field decay steeper than the
ideal-MHD case, can be taken as a phenomenological description of a
resistive regime. Under these conditions, in fact, a finite electrical
conductivity will lead to a more rapid decay of the magnetic field and,
in turn, to a slower decay of the fluid energy density, which is
``heated up'' by the decaying field [\cf Eq.\  \eqref{eq:general_soln_4}].

Let us start by considering the equation of energy conservation
\eqref{eq:1dMHD}, which for a general EOS of the form
(\ref{eq:EOS}) and a magnetic-field evolution given by Eq.\
\eqref{eq:BasTau} yields
\begin{equation}
\partial_{\tau}\left[\ \tilde{e}
  +\frac{\sigma_{0}}{2}\left(\frac{\tau_{0}}{\tau}\right)^{2a}\right]+\left(
1+{c}_{s}^{2} \right) \frac{\tilde{e}}{\tau}+
\frac{\sigma_{0}}{\tau}\left(\frac{\tau_{0}}{\tau}\right)^{2a}=0\,.
\label{eq:1dMHD_Case2_02}
\end{equation}
It is not difficult to find the analytic solution of this equation
with initial condition $\tilde{e}_{0} = 1$,
\begin{equation}
\label{eq:general_soln_4}
\tilde{{e}}(\tau) ={\left( \frac{\tau_{0}}{\tau} \right)}^{1+ c_{s}^{2}}
+{\sigma}_{0}\frac{1-a}{1+c_s^2-2a} \left[ \left(
    \frac{\tau_{0}}{\tau} \right)^{1+ c_{s}^{2}}-\left(
    \frac{\tau_{0}}{\tau} \right)^{2a} \right]\,.
\end{equation}

Once again, it is possible to see that in the limit of vanishing
magnetization $\sigma_{0}\rightarrow0$, Eq.\ (\ref{eq:general_soln_4})
coincides with the solution \eqref{eq:analytic_withB_s} for Bjorken
flow. Furthermore, for $\sigma_0 \neq 0$ but $a=1$, the solution
(\ref{eq:general_soln_4}) also coincides with
Eq.\ \eqref{eq:analytic_withB_s}, thus highlighting that
Eq.\ \eqref{eq:BasTau} with $a=1$ represents the evolution equation for a
magnetic field in the ideal-MHD limit.

Note that, for the second term in Eq.\ \eqref{eq:general_soln_4}, the
sign of the expression in brackets divided by $1+c_s^2-2a$ is always
negative (remember that $c_s^2\leq 1$ by causality). Thus, for the case
$a>1$ the second term is always positive. As a result, it always leads to
a slower decay (and sometimes, as we will show below, even to an
intermittent increase) of the fluid energy density than in the case $a
=1$.  Viceversa, for $a<1$ the second term in
Eq.\ \eqref{eq:general_soln_4} is always negative, leading to a faster
decay than in the case $a=1$.

Equation (\ref{eq:general_soln_4}) seems to have a divergent behaviour at
$a=(1+c_s^2)/2$, but this is only a first impression. 
We demonstrate in App.\ \ref{appen:Limit} that
in the limit $a\rightarrow(1+c_s^2)/2$,
\begin{equation}
\lim_{a\rightarrow (1+c_s^2)/2} \frac{ \left( \tau_{0}/\tau
      \right)^{1+c_s^2}-\left( \tau_{0}/\tau \right)^{2a} }{ 1+c_s^2-2a} 
      = \left( \frac{\tau_{0}}{\tau} \right)^{1+c_s^2}\ln
     \left(\frac{\tau_{0}}{\tau}\right)\,,
\end{equation}
Collecting things, the final solution of the energy-conservation equation
(\ref{eq:1dMHD_Case2_02}) for $a = (1+c_s^2)/2$ is
\begin{equation}
\label{eq:general_soln_2}
\tilde{e}(\tau)
=\left(\frac{\tau_0}{\tau}\right)^{1+c_s^2}+
\sigma_{0}\, \frac{1-c_s^2}{2}\left(\frac{\tau_{0}}{\tau}\right)^{1+c_s^2}
\ln\left(\frac{\tau_{0}}{\tau}\right)\,.
\end{equation}
Note that for $\tau \geq \tau_0$ the second term on the right-hand side
of Eq.\  \eqref{eq:general_soln_2} is negative, hence increasing the decay
of the fluid energy density with respect to the $a=1$ case. 
Furthermore, while in the ideal-MHD case the
solution $\tilde{e}=0$ is obtained only asymptotically, for $a=(1+c_s^2)/2$ this
extreme case is obtained after a finite time $\bar{\tau}=\tau_0
e^{2/[(1-c_s^2)\sigma_0]}$.

In summary, Eqs.\  \eqref{eq:general_soln_4} and \eqref{eq:general_soln_2}
represent the solutions to Eq.\  (\ref{eq:1dMHD_Case2_02}); furthermore, since
Eq.\  \eqref{eq:general_soln_4} comprises also the case $a=1$,
these equations provide a rather complete description of 
the full solution to the energy-conservation equation
(\ref{eq:1dMHD_Case2_02}). As an example, we quote the solutions for $c_s^2=1/3$.
For $a\neq 2/3$,
\begin{equation}
\label{eq:general_soln_5}
\tilde{{e}}(\tau) ={\left( \frac{\tau_{0}}{\tau} \right)}^{4/3}
+\frac{\sigma_{0}}{2}\,\frac{1-a}{2/3-a} \left[ \left(
    \frac{\tau_{0}}{\tau} \right)^{4/3}-\left(
    \frac{\tau_{0}}{\tau} \right)^{2a} \right]\,,
\end{equation}
and for $a=2/3$,
\begin{equation}
\label{eq:general_soln_6}
\tilde{e}(\tau)
=\left(\frac{\tau_0}{\tau}\right)^{4/3}+
\frac{\sigma_{0}}{3}\left(\frac{\tau_{0}}{\tau}\right)^{4/3}
\ln\left(\frac{\tau_{0}}{\tau}\right)\,.
\end{equation}

\section{Discussion}
\label{sec:results}

\begin{figure}
\includegraphics[width=0.8\columnwidth]{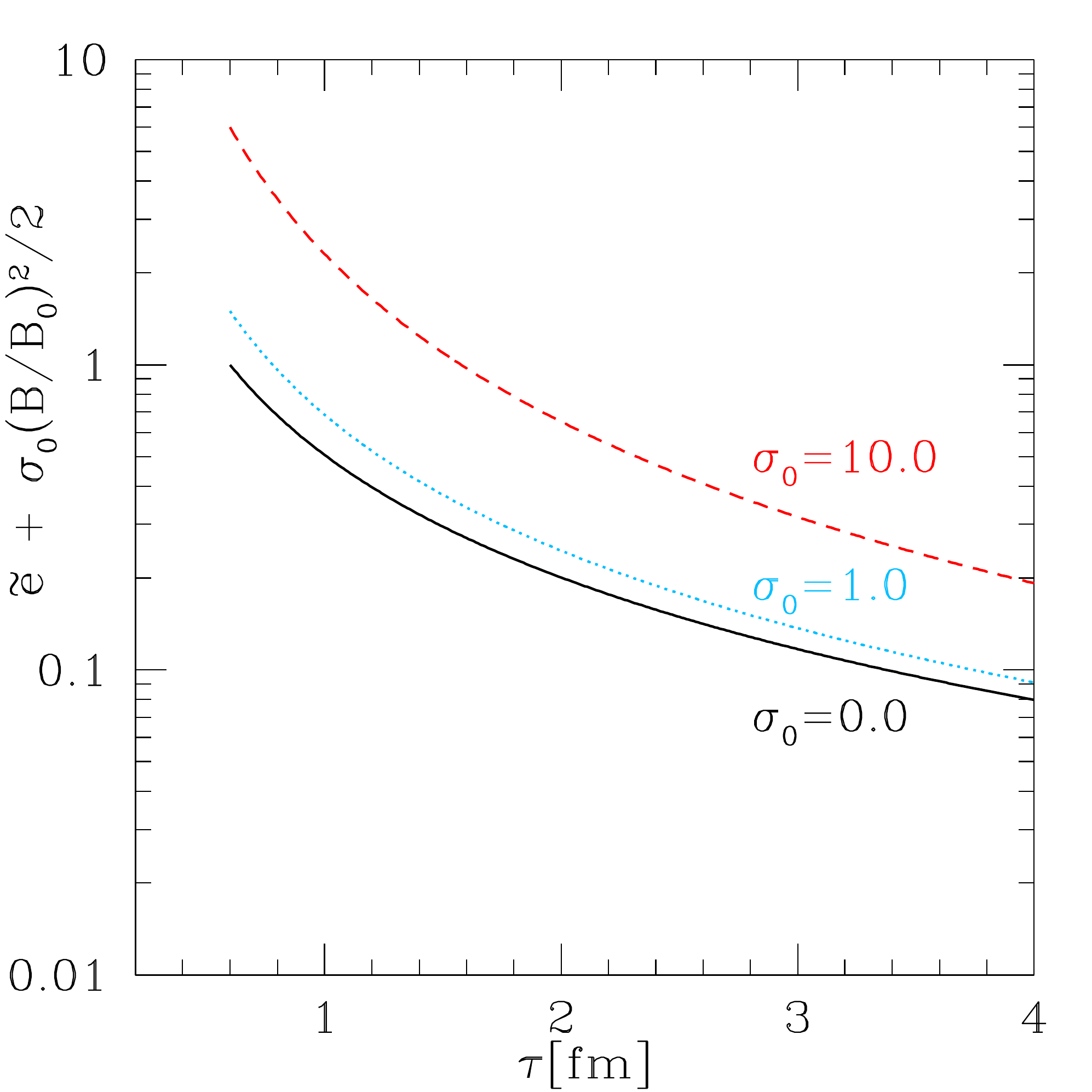}
\caption{Evolution of the normalised total energy density $\tilde{e} +
  \tfrac{1}{2}{\sigma_{0}}(B/B_0)^2$. Different lines refer to different
  values of the initial magnetization, ranging from $\sigma_{0}=0$ (solid
  black line) up to cases with initial magnetization of $\sigma_{0}=1$
  (light-blue dotted line) and $\sigma_0=10$ (red dashed line). Note that
  the fluid energy density decays like $\tau^{-4/3}$ for all values of
  $\sigma_0$, \ie as in traditional Bjorken flow.}
\label{eps_bjorken_1d}
\end{figure}

This section is devoted to a discussion of the various analytic solutions
found in the previous section. For the sake of definiteness, we will
always use the value $c_s^2=1/3$. We start by considering the ideal-MHD
case, in which case the time evolution of the energy density and the
magnetic field is given by Eqs.\ \eqref{eq:analytic_withB_s} and
\eqref{eq:Bs_evol}, respectively. These solutions are shown in
Fig.\ \ref{eps_bjorken_1d} which reports the evolution of the normalised
total energy density $\tilde{e} + \tfrac{1}{2}{\sigma_{0}}(B/B_0)^2$ for
$\tau_0=0.6\,{\rm fm}$. Different lines refer to different values of the
initial magnetization, ranging from $\sigma_{0}=0$ (Bjorken flow without
magnetic field; black solid line) up to cases with initial magnetization
of $\sigma_{0}=1$ (light-blue dotted line) and $\sigma_0=10$ (red dashed
line). As already discussed in the previous section, the evolution of the
fluid energy density does not depend on $\sigma_0$ [\cf
  Eq.\ \eqref{eq:analytic_withB_s}] and scales like $\tau^{-4/3}$, while
the magnetic energy density scales like $\tau^{-2}$. As a result,
increasing $\sigma_0$ (as we do in Fig.\ \ref{eps_bjorken_1d}) only adds
energy density to the system, but does not alter the temporal evolution
of the fluid energy density.

Having considered the simple case $a=1$, we next discuss the behaviour of
the solutions when the magnetic field varies according to the more
general power law (\ref{eq:BasTau}). We have already mentioned that $a >
1$ corresponds to the case when the magnetic field decays faster than in
the ideal-MHD limit and could therefore be associated to a resistive
regime. Conversely, a magnetic-field evolution with $a < 1$ would
correspond to a decay that is slower than in the ideal-MHD limit. As the
case $a=1$ is the limit of infinite conductivity, and thus of a maximal
magnetic induction, it is at first sight hard to imagine how to produce a
magnetic field that decays even slower than in the ideal-MHD
case. However, in heavy-ion collisions the remnants of the colliding
nucleons can give an additional contribution to the magnetic field,
slowing down its decay \cite{Deng:2012pc}. Thus, considering also the
case $a <1$ is reasonable in this context. Within this range, a
particularly interesting solution is the one where $a=2/3$, for in this
case the general solution \eqref{eq:general_soln_5} needs to be replaced
by the special solution \eqref{eq:general_soln_6}.

\begin{figure}
\includegraphics[width=0.8\columnwidth]{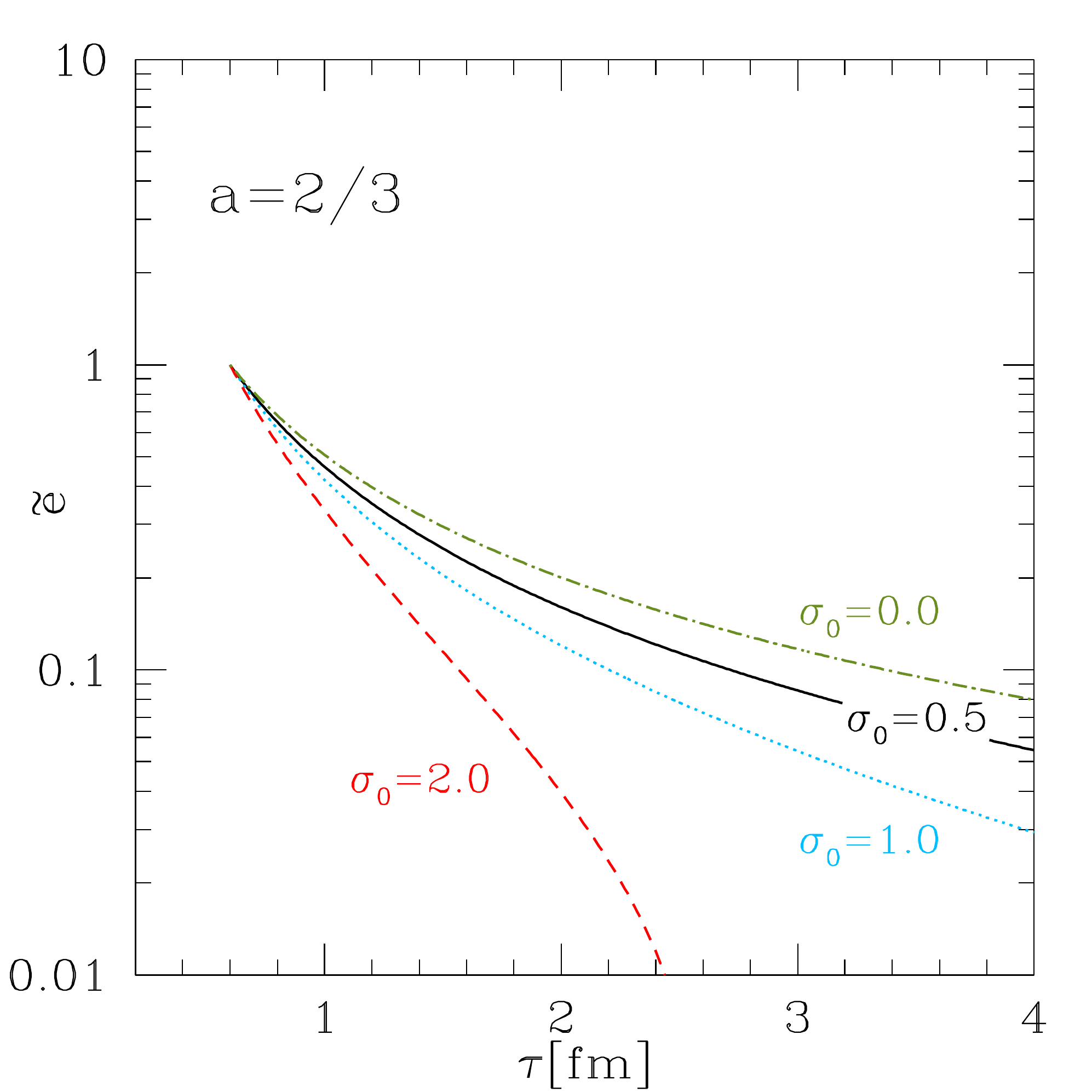}
\caption{Evolution of the normalized fluid energy density $\tilde{e}$ for
  $a={2}/{3}$. Different lines refer to different levels of the initial
  magnetization: $\sigma_{0}= 0.5$ (black solid line), $\sigma_0=1$
  (dotted light-blue line), and $\sigma_0=2.0$ (red dashed line). Note
  that the decrease of the energy density is always faster than in the
  ideal-MHD case (green dot-dashed curve).}
\label{fig:a2b3} 
\end{figure}

\begin{figure}
\includegraphics[width=0.8\columnwidth]{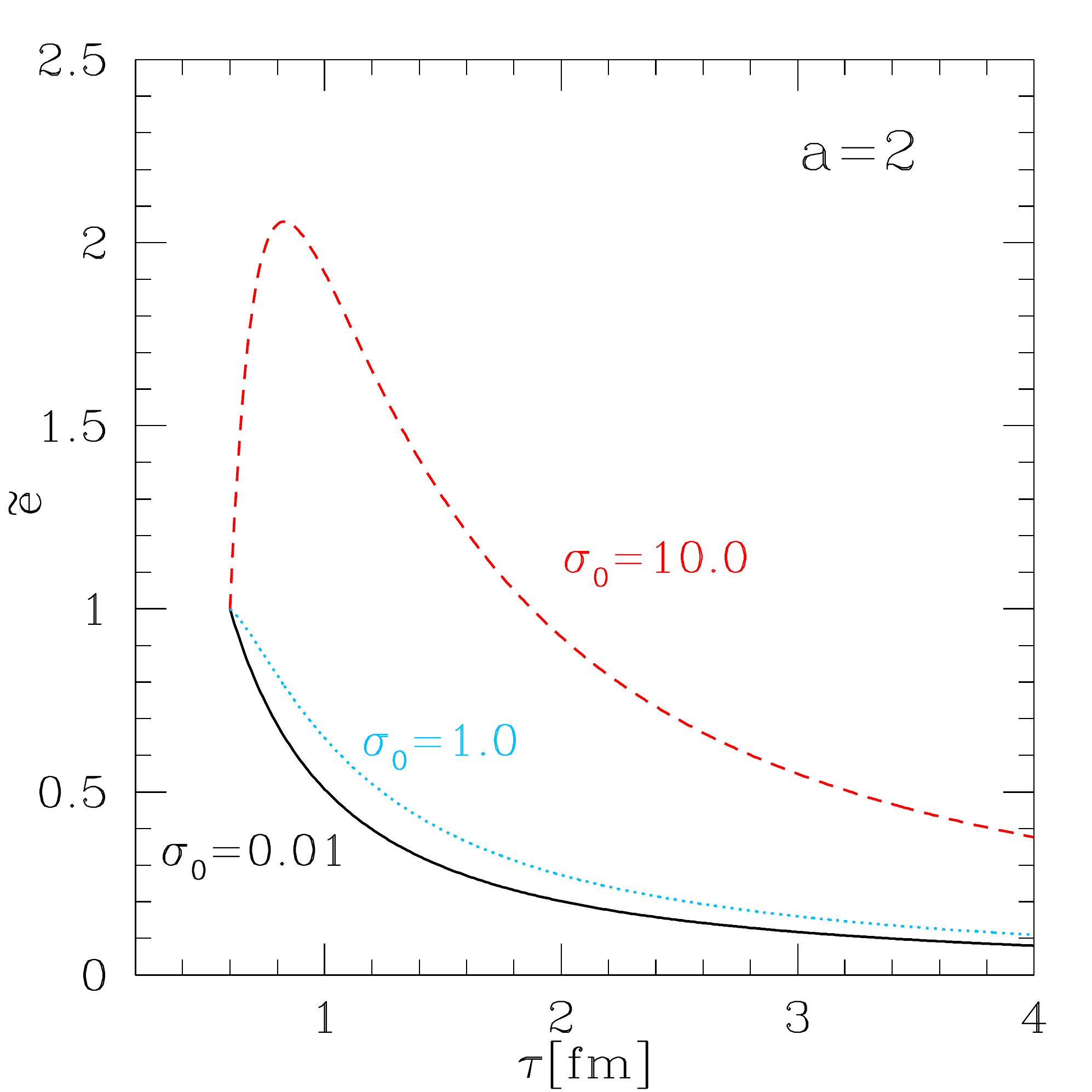}
\caption{Evolution of normalized fluid energy density $\tilde{e}$ for a
  magnetic field with a power-law decay $a = 2$. Also in this case,
  different lines refer to different levels of the initial magnetization,
  ranging from $\sigma_{0}= 0.01$ (black solid line), $\sigma_0=1$
  (light-blue dotted line), and $\sigma_0=10$ (red dashed line). Note the
  initial ``heat-up'' in the case of large magnetizations.}
\label{fig:a50} 
\end{figure}

\begin{figure}
\includegraphics[width=0.8\columnwidth]{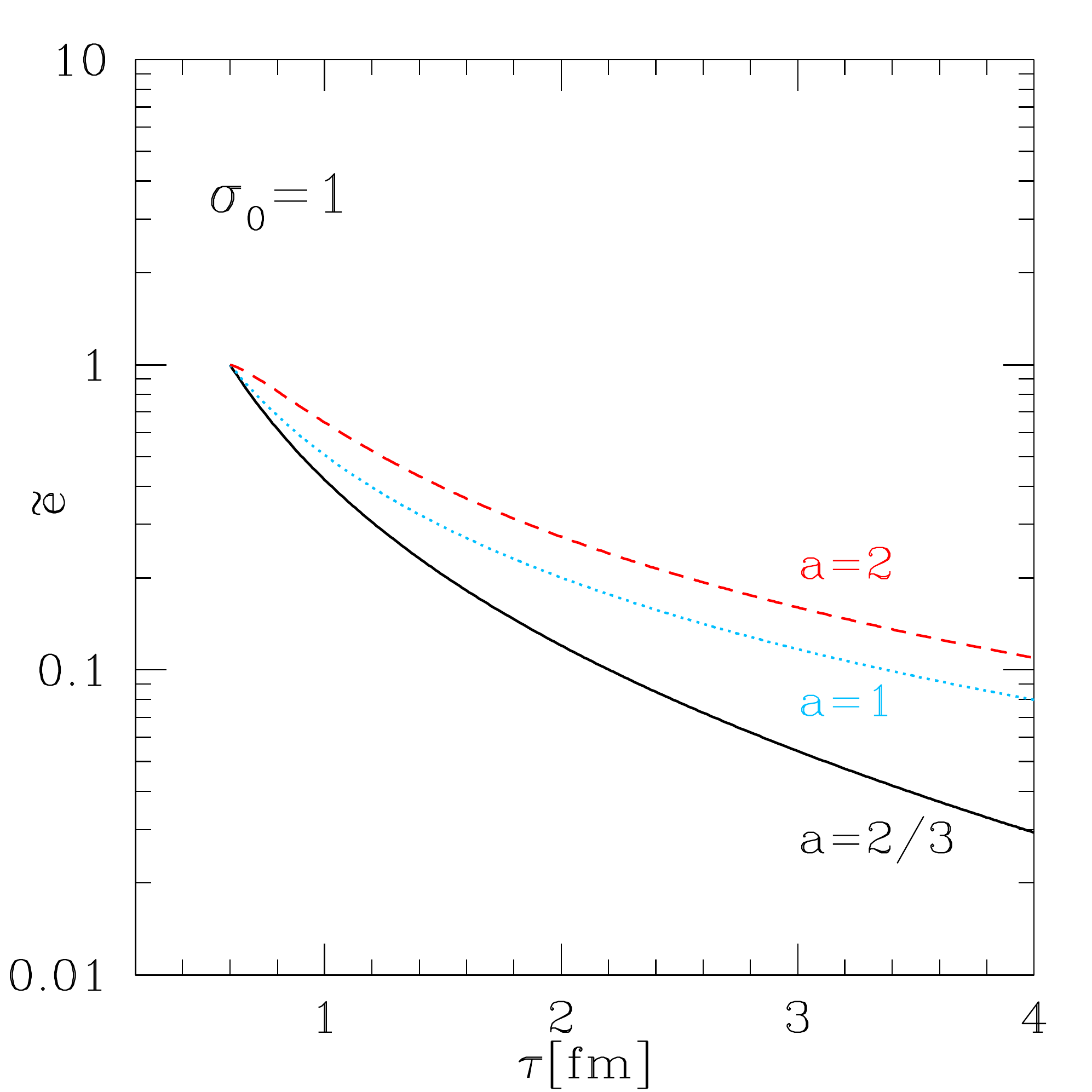}
\caption{Evolution of the normalized energy density $\tilde{e}$ in the
  different cases and when the initial magnetization is set to
  $\sigma_0=1$. Different lines refer to the evolution for $a= 2/3$
  (black solid line), $a=1$ (light-blue dotted line), and $a=2$ (red
  dashed line). Clearly, $\tilde{e}$ decreases more rapidly for $a=2/3$
  when compared to the case $a=1$, whereas for $a=2$ it initially
  decays more slowly and then decays asymptotically at the same rate as
  for the ideal-MHD case $a=1$.}
\label{fig:Bjorken_three_a} 
\end{figure}

Such a solution is shown in Fig.\ \ref{fig:a2b3}, which reports the
evolution of $\tilde{e}$ for $a={2}/{3}$ and where different lines refer
to different levels of the initial magnetization: $\sigma_{0}= 0.5$
(black solid line), $\sigma_0=1$ (dotted light-blue line), and
$\sigma_0=2.0$ (red dashed line). As already anticipated in the previous
section, for $\tau>\tau_0$ the log term always reduces the value of
$\tilde{e}$, leading to a faster decrease of the energy density when
compared with the ideal-MHD limit (this is shown with a green dot-dashed
line in Fig. \ref{fig:a2b3}). Furthermore, it is also clear that larger
values of $\sigma_{0}$ will lead to a faster decrease in $\tilde{e}$, as
shown in Fig.\ \ref{fig:a2b3} (note that for $\sigma_{0}=2$,
$\tilde{e}=0$ at $\tau \simeq 2.7\,{\rm fm}$).

Finally, we consider in Fig.\ \ref{fig:a50} the evolution of normalized
fluid energy density $\tilde{e}$ in the case $a = 2$. Also in this case,
different lines refer to different levels of the initial magnetization,
$\sigma_{0}= 0.01$ (black solid line), $\sigma_0=1$ (light-blue dotted
line), and $\sigma_0=10$ (red dashed line). Because the second term in
Eq.\ \eqref{eq:general_soln_5} is always positive, the evolution of the
energy density is expected to be slower than in standard Bjorken flow
(see also Fig.\ \ref{fig:Bjorken_three_a}). At the same time, the second
term in Eq.\ \eqref{eq:general_soln_5} is not a monotonically decreasing
function of $\tau$. As a result, it may produce even a temporary increase
in the fluid energy density. This increase, which can be associated with
a resistive ``heating up'' of the fluid, will depend on the values of
$\sigma_0$ and $a$ and will be larger for larger values of the
latter. This is clearly shown in Fig.\ \ref{fig:a50} for $\tau \lesssim
1\,{\rm fm}$ in the case $\sigma_{0}=10.0$; after this time the evolution
of the energy density is monotonically decreasing and asymptotically
dominated by the term $\sim \tau^{-4/3}$.

As a way to summarize the various results presented so far we show in
Fig.\ \ref{fig:Bjorken_three_a} the evolution of the normalized energy
density $\tilde{e}$ in the different cases but keeping the
initial magnetization fixed to $\sigma_0=1$. More specifically, we show the
evolution for $a= 2/3$ (black solid line), $a=1$ (light-blue dotted
line), and $a=2$ (red dashed line). Clearly, $\tilde{e}$ decreases more
rapidly for $a=2/3$ when compared to the case $a=1$, whereas for
$a=2$ it initially decreases more slowly and then decays asymptotically
at the same rate as for the ideal-MHD $a=1$ case.

\section{Conclusions}
\label{sec:summary}

Driven by the interest in exploring the effects of strong magnetic fields
in the hydrodynamical description of relativistic heavy-ion collisions,
we have studied the evolution of the fluid energy density following the instant
of the collision and considering an ultrarelativistic fluid with EOS
$p=c^2_s\,e$. Because we are mainly interested in finding analytic
solutions, our setup is somewhat idealized and we have therefore
considered one-dimensional, longitudinally boost-invariant
flow with transverse magnetic field,
\ie a transverse-MHD flow. When no magnetic fields are present, this flow
is known as the Bjorken flow \cite{Bjorken:1982qr} and although it
represents a simplified prescription, it has served to gain significant
insight on the dynamics of nucleus-nucleus collisions.

We have first considered the dynamics of a one-dimensional MHD flow in
the limit of infinite electrical conductivity and found a somewhat
surprising result, namely that in the ideal-MHD limit the Bjorken flow applies
unmodified. The evolution of the fluid energy density, in fact, is regulated by
the same equation found for Bjorken flow and thus with an analytic
decay in proper time as $\tau^{-4/3}$. Of course large values of the
initial magnetization will change the values of the total energy density,
but the evolution of the fluid energy density will not be modified
because of the passive role played by the magnetic field in this
regime. This result widens considerably the range of applicability of the
Bjorken model and shows that it can be used, unmodified, also to describe
collisions in transverse MHD.

We have also considered the cases in which the magnetic-field evolution
is not the one prescribed by the ideal-MHD limit but, rather, follows a
power-law behaviour in proper time with exponent $a$. The solutions in
this case need to be distinguished between the scenario in which the
magnetic field decays more slowly than in the ideal-MHD case, \ie for $a
< 1$ and when the decay is more rapid, \ie for $a > 1$. In the first
scenario, which could be realized when remnants of the colliding nuclei
slow down the decay of the magnetic field, the decay of the energy density is 
faster. Furthermore, the rate at which this decay takes place is
determined entirely by the level of the initial magnetization and modeled
in terms of the dimensionless magnetic-to-fluid energy $\sigma_0 \equiv
B^2_0/e_0$. In the second scenario, which could be associated to
a resistive regime in which magnetic field energy is converted to
fluid energy via resistive losses, the evolution of the energy density
is more complex. In the initial stages of the evolution, in fact, the fluid
energy density may increase as it would in terms of a resistive ``heating
up'' of the fluid. The amount of this increase depends on the magnetic
field strength and dissipation and hence will increase with $\sigma_0$
and $a$. However, as the fluid further expands, its energy density will
decrease with an asymptotic rate that is the same as in the Bjorken flow,
\ie $\propto \tau^{-4/3}$.

The work presented here could be extended in a number of ways. First, one could
search for analytic solutions in one-dimensional MHD in a Landau-type 
flow scenario. Second, one can consider an explicitly finite electrical conductivity
as the simplest model for a one-dimensional MHD flow with chiral
fermions. Results on these topics will be presented in forthcoming
papers.

\begin{acknowledgments}
V.R.\ and S.P.\ are supported by the Alexander von Humboldt Foundation,
Germany.  Partial support comes from "NewCompStar", COST Action MP1304 
and the NSFC under grant No. 11205150.
\end{acknowledgments}

\appendix


\section{Covariant derivative for Bjorken expansion}
\label{Appendix2}

In this appendix we sketch briefly the steps that are needed to derive
the energy-and momentum-conservation equations discussed in the main
text. Before we start, we should comment about the notation used in
  defining the energy-momentum tensor \eqref{eq:EMTensor}. Bearing in
  mind that in general-relativistic calculations the standard choice for
  the signature is a spacelike one, \ie $(-,+,+,+)$, the energy-momentum
  tensor in GRMHD is normally defined as
  \cite{Giacomazzo:2005jy,Giacomazzo:2007ti}
\begin{eqnarray}
\label{eq:EMTensorGRMHD}
{T}^{\mu\nu}=\left(e+p+{b}^{2}\right){u}^{\mu}{u}^{\nu}+
\left(p+\frac{{b}^{2}}{2}\right){g}^{\mu\nu}-{b}^{\mu}{b}^{\nu}\,,
\end{eqnarray}
where $\boldsymbol{b}$ is the magnetic field four-vector measured in a
comoving frame and has components given by
\begin{eqnarray}
{b}^{\mu}=\left(\gamma\; 
\vec{\boldsymbol{v}}\cdot\vec{\boldsymbol{B}},
\frac{\vec{\boldsymbol{B}}}{\gamma}+
\gamma\;\vec{\boldsymbol{v}}\cdot\vec{\boldsymbol{B}}\;
\vec{\boldsymbol{v}}\right)\,,
\end{eqnarray}
with $\gamma \equiv 1/{\sqrt{1-v^{2}}}$ the the Lorentz factor and
$\vec{\boldsymbol{B}}$ is the magnetic field three-vector measured by an
Eulerian (or normal) observer. The modulus of ${b}^{\mu}$ is then given
by
\begin{eqnarray}
{b}^{2}\equiv{b}^{\mu}{b}_{\mu}=\frac{\vec{\boldsymbol{B}}^{2}}{{\gamma}^{2}}+
{\left(\vec{\boldsymbol{v}}\cdot\vec{\boldsymbol{B}}\right)}^{2}\,.
\end{eqnarray}
With this clarification in mind, we go back to our special-relativistic
setting with energy-momentum \eqref{eq:EMTensor} and consider the
projection of the energy-momentum conservation equation
${\partial}_{\nu}{T}^{\mu\nu}=0$ along the fluid four-velocity $u^{\mu}$,
which reads
\begin{widetext}
\begin{eqnarray}
{u}_{\mu}{\partial}_{\nu}{T}^{\mu\nu} &=& 0\\
%
%
{u}_{\mu}{{u}^{\mu}{u}^{\nu}\partial}_{\nu}\left(e + p + {B}^{2}\right) +
\left(e + p +
     {B}^{2}\right){u}_{\mu}{\partial}_{\nu}\left({u}^{\mu}{u}^{\nu}\right)
     - {u}_{\mu}{\partial}_{\nu}\left[ \left(p +
       \frac{{B}^{2}}{2}\right){g}^{\mu\nu}\right] -
     {u}_{\mu}{\partial}_{\nu}\left(B^{\mu}B^{\nu}\right) &=& 0\,,\nonumber \\
{{u}^{\nu}\partial}_{\nu}\left(e + p + {B}^{2}\right) + \left(e + p +
{B}^{2}\right){\partial}_{\nu}{u}^{\nu}-{u}^{\nu}{\partial}_{\nu}\left(p
+ \frac{{B}^{2}}{2}\right) +
B^{\mu}B^{\nu}{\partial}_{\nu} {u}_{\mu} &=& 0\,,\nonumber \\
\partial_{\tau}\left(e + p + {B}^{2}\right) + \frac{e + p +
  {B}^{2}}{\tau} - \partial_{\tau}\left(p +
\frac{{B}^{2}}{2}\right) &=& 0\,,\nonumber
\\ \partial_{\tau}\left(e + \frac{{B}^{2}}{2}\right) +
\frac{e + p + {B}^{2}}{\tau} &=& 0\,,\nonumber
\end{eqnarray}
\end{widetext}
where we have used Eq.\ (\ref{eq:EMTensor}), 
$u^{\mu}B_{\mu}=0$ and $B^{\nu} \partial_{\nu} u_\mu=0$, 
since $u_{\mu} = \left(u_0,0,0,u_z\right)$ and $B_{\mu} =
\left(0,B_x,B_y,0\right)$ in our transverse-MHD setup.

Similarly, the projection of the conservation equation
${\partial}_{\nu}{T}^{\mu\nu}=0$ in the direction orthogonal to the fluid
four-velocity gives
\begin{widetext}
\begin{eqnarray}
h_{\mu\nu}\partial_{\alpha}T^{\alpha\nu} & = & 0\,,\\
(e + p + B^{2})h_{\mu\nu}\partial_{\alpha}(u^{\alpha}u^{\nu}) -
h_{\mu\nu}\partial^{\nu}\left(p + \frac{B^{2}}{2}\right) -
h_{\mu\nu}\partial_{\alpha}(B^{\alpha}B^{\nu}) & = & 0\,,\nonumber \\
%
%
(e + p + B^{2})u^{\alpha}\partial_{\alpha}u_{\mu} -
h_{\mu\nu}\partial^{\nu}\left(p + \frac{B^{2}}{2}\right) -
B^{\alpha}\partial_{\alpha}B_{\mu} - B_{\mu}\partial^{\alpha}B_{\alpha} +
u_{\mu}u_{\nu}\partial_{\alpha}\left(B^{\alpha}B^{\nu}\right) & = & 0\,, \nonumber\\
(e + p + B^{2})\partial_{\tau}u_{\mu} -
h_{\mu\nu}\partial^{\nu}\left(p + \frac{B^{2}}{2}\right) -
B^{\alpha}\partial_{\alpha}B_{\mu} - B_{\mu}\partial^{\alpha}B_{\alpha} -
u_{\mu}B^{\alpha}B^{\nu}\partial_{\alpha}u_{\nu} & = & 0\,, \nonumber
\\ \nonumber
\end{eqnarray}

\end{widetext}
where we have introduced $\boldsymbol{h}$ as the orthogonal projector to
$\boldsymbol{u}$, \ie $\boldsymbol{h} \cdot \boldsymbol{u} = 0$, where
$h_{\mu\alpha} \equiv \eta_{\mu \nu} - u_{\mu}u_{\nu}$.
The last three terms vanish, because $B_\mu$ is assumed to be
constant in transverse direction. This then leads to Eq.\ (\ref{eq:Euler}).


\section{Frozen-flux theorem}
\label{appen:frozenFlux}

In this appendix we show that the evolution of the magnetic field and of
the entropy density are strictly related in the ideal-MHD limit. The
arguments reported below are well known and can be found in a number of
textbooks (\eg Refs.\ \cite{Landau:1987,Rezzolla:2013}), but we recall them here
for completeness. We start from the definition of the covariant (or
Lagrangian or convective) time derivative given by
\begin{equation}
\frac{D}{Dt}\equiv\frac{\partial}{\partial t} +
\vec{\boldsymbol{u}}\cdot\vec{\nabla}\,,
\end{equation}
where $\vec{\boldsymbol{u}}$ is the fluid velocity. If $\vec{\boldsymbol{x}}$
is the position of a fluid element, this will be advected with the flow
and hence have 
\begin{equation}
\frac{D \vec{\boldsymbol{x}}}{Dt} = 0\,.
\end{equation}
However, if $\vec{\boldsymbol{\xi}}$ is a vector separating two fluid elements at a given
instant, the corresponding Lagrangian derivative will
not be necessarily be zero, but is actually given by
\begin{equation}
\frac{D\vec{\boldsymbol{\xi}}}{Dt} =
\vec{\boldsymbol{\xi}}\cdot\vec{\nabla}\vec{\boldsymbol{u}}\,.
\label{eq:material_advection}
\end{equation}
Stated differently ${D}\vec{\boldsymbol{\xi}}/{Dt} = 0$ only for a fluid
with uniform velocity $\vec{\boldsymbol{u}}$. In all other cases, the
vector $\vec{\boldsymbol{\xi}}$ will change its length and/or orientation
in the presence of a velocity gradient.

Next, we consider the conservation of rest mass, which can be expressed
as

\begin{equation}
\frac{D\rho}{Dt}=-\rho\vec{\nabla}\cdot\vec{\boldsymbol{u}},
\end{equation}
where $\rho$ is the rest-mass density of the fluid. In ideal MHD the
induction equation takes the well-known form
\begin{equation}
\frac{\partial\vec{\boldsymbol{B}}}{\partial t}=\vec{\nabla}\times
\left(\vec{\boldsymbol{u}}\times\vec{\boldsymbol{B}}\right)\,,
\label{eq:magnetic_induction}
\end{equation}
and the frozen-flux theorem states that the magnetic field lines are
frozen in the fluid and can be identified with the worldlines of fluid
elements. To see this, we use the following vector identity
\begin{equation}
\vec{\nabla}\times\left(\vec{\boldsymbol{u}}\times\vec{\boldsymbol{B}}\right)=
\vec{\boldsymbol{B}}\cdot\vec{\nabla}\vec{\boldsymbol{u}} -
\vec{\boldsymbol{B}}\left(\vec{\nabla}\cdot\vec{\boldsymbol{u}}\right)-
\vec{\boldsymbol{u}}\cdot\vec{\nabla}\vec{\boldsymbol{B}} +
\vec{\boldsymbol{u}}\left(\vec{\nabla}\cdot\vec{\boldsymbol{B}}\right)\,,
\end{equation}
together with 
$\vec{\nabla}\cdot\vec{\boldsymbol{B}}=0$ in
Eq.\  \eqref{eq:magnetic_induction} to obtain
\begin{equation}
\frac{D\vec{\boldsymbol{B}}}{Dt} =
\vec{\boldsymbol{B}}\cdot\vec{\nabla}\vec{\boldsymbol{u}} -
\vec{\boldsymbol{B}}\left(\vec{\nabla}\cdot\vec{\boldsymbol{u}}\right)\,.
\label{eq:B6}
\end{equation}
Together with the conservation of mass, the above equation can then be
written as
\begin{equation}
\frac{D}{Dt}\left(\frac{\vec{\boldsymbol{B}}}{\rho}\right)=
\frac{\vec{\boldsymbol{B}}}{\rho}
\cdot\vec{\nabla}\vec{\boldsymbol{u}}\,,
\label{eq:FrozenFluxRho}
\end{equation}

This is exactly the same equation satisfied by the separating vector
$\vec{\boldsymbol{\xi}}$ [Eq.\  \eqref{eq:material_advection}]. Therefore
a magnetic field line is advected and distorted by the fluid in the same
way as a fluid element. If the fluid expansion takes place
isentropically, the total entropy of the system remains constant and the
entropy density $s$ will satisfy the same conservation equation as the
rest-mass density, \ie
\begin{equation}
\frac{Ds}{Dt}=-s\vec{\nabla}\cdot\vec{\boldsymbol{u}}\,.
\label{eq:EntropyCon}
\end{equation}

From the arguments made above, it follows that the quantity
$\vec{\boldsymbol{B}}/{s}$ will behave as the quantity
$\vec{\boldsymbol{B}}/{\rho}$ and hence satisfy the equation
\begin{equation}
\frac{D}{Dt}\left(\frac{\vec{\boldsymbol{B}}}{s}\right)=
\frac{\vec{\boldsymbol{B}}}{s}\cdot\vec{\nabla}
\vec{\boldsymbol{u}}\,,
\end{equation}
which is identical with Eq.\ \eqref{eq:FrozenFluxRho} except $\rho$ is
replaced by $s$, \ie for this case we also have the magnetic fluxes
frozen in the system.

\section{Limit for the log term}
\label{appen:Limit}

In this appendix we discuss how to evaluate the second term in
Eq.\  \eqref{eq:general_soln_5} in the limit in which $a\to (1+c_s^2)/2$, \ie
the limit
\begin{equation}
\label{eq:lim_1}
\lim_{a\rightarrow (1+c_s^2)/2} {\frac{ \left( \tau_{0}/\tau
      \right)^{1+c_s^2}-\left( \tau_{0}/\tau \right)^{2a} }{1+c_s^2-2a}}\,.
\end{equation}
We first increase the power exponent $a$ by an infinitesimal amount
$\epsilon>0$ and then take the limit $\epsilon\rightarrow 0 $. In this
case, Eq.\  \eqref{eq:lim_1} becomes
\begin{equation}
\lim_{\epsilon \rightarrow 0} \frac{ \left( 
\tau_{0}/\tau \right)^{1+c_s^2}-\left( 
\tau_{0}/\tau \right)^{2\left( a + \epsilon \right)} }{
    1+c_s^2 -2\left( a+\epsilon \right) }\,,
\end{equation}
and setting $a=(1+c_s^2)/2$ we obtain the desired result
\begin{eqnarray}
\nonumber \left(\frac{ \tau_{0}}{\tau} \right)^{1+c_s^2}
\lim_{\epsilon \rightarrow  0}
{\frac{ 1-\left( \tau_{0}/\tau
      \right)^{2\epsilon} }{\left( -2\epsilon \right)}} = 
\left(\frac{\tau_{0}}{\tau} \right)^{1+c_s^2}\ln \left(\frac{\tau_{0}}{\tau}\,\right).
\end{eqnarray}

\end{document}